\newcommand{\bea}{\begin{eqnarray}}
\newcommand{\eea}{\end{eqnarray}}
\newcommand{\bear}{\begin{eqnarray*}}
\newcommand{\eear}{\end{eqnarray*}}

\def\ffrac#1#2{\textstyle{#1\over#2}\displaystyle}

\documentstyle[sprocl]{article}
 
\input{psfig}

\begin{document}


\title
{AN EXACTLY SOLVABLE CONSTRAINED XXZ CHAIN$^*$}

\author{
F.C. ALCARAZ$^1$ and R.Z. BARIEV$^{1,2}$}

\address{$^1$Departamento de F\'{\i}sica, 
Universidade Federal de S\~ao Carlos,\\ 
13565-905, S\~ao Carlos, SP Brazil}

\address{$^2$The Kazan Physico-Technical Institute of the Russian 
Academy of Sciences,\\ Kazan 420029, Russia}

\maketitle\abstracts{
A new family of exactly solvable models is introduced. These models 
are generalizations of the XXZ chain where the distance among
spins up ($\sigma^z$-basis) cannot be smaller or equal to 
$t$ $(t=0,1,2,...)$. The case $t=0$ recovers the standard XXZ 
chain. The coordinate Bethe ansatz is applied and the phase 
diagram is calculated. Exploring the 
 finite-size consequences of conformal 
invariance the critical exponents are evaluated exactly at the 
critical regions of the 
  phase diagram.}


\footnote{To appear in the Proceedings of the Conference: Statistical 
Physics on the Eve of the Twenty-Fist Century}

\section{Introduction}

The anisotropic $S=\frac{1}{2}$ Heisenberg model, or XXZ chain, is one 
of the most studied quantum spin system in statistical mechanics. Since 
its exact solution by Yang and Yang\,\cite{1} this model has been 
considered to be the classic example of the success of the Bethe ansatz. 
In this paper, considering the XXZ chain as a prototype we introduce a 
new set of quantum chains which preserve the property of integrability.

The XXZ Hamiltonian on an $L$-site chain is given by
\bea
H(\Delta) = -\frac{1}{2}\sum_{i=1}^L(\sigma_i^x\sigma_{i+1}^x +
\sigma_i^y\sigma_{i+1}^y + \Delta\sigma_i^z\sigma_{i+1}^z) \,, 
\eea
where $\sigma^x,\sigma^y,\sigma^z$ are spin-$\frac{1}{2}$ Pauli matrices, 
$\Delta$ is the anisotropy parameter 
and we assume periodic boundary conditions.
If we identify, in the $\sigma^z$-basis, a spin up or spin
down at a given site $i$ as a particle or hole at this site,
the XXZ Hamiltonian can be written iquivalently as
\bea
H(\Delta) = - \sum_{i+1}^L P_{0}(c_i^+c_{i+1} + c_{i+1}^+c_i + 
\Delta n_in_{i+1})P_0 \,,
\eea
where $c_i^+$ and $c_i$ are spinless creation and annihilation
operators and $n_i = c_i^+c_i$ is the density of particles at site $i$. 
The projector $P_o$ in (1) forbids two particles to occupy the same 
position in space (fermion like). The models under consideration are 
obtained by a generalization of the operator $P_0$ and are given by
\bea
H(\Delta) = - \sum_{i=1}^L P_t(c_i^+c_{i+1} + c_{i+1}^+c_i + 
\Delta n_in_{i+ t +1})P_t\,, \hspace{1cm} t = 0,1,2,...
\eea
where  $P_t$ now forbids two particles at distance less or equal to 
$t$ $(0,1,2,...)$. The static interaction,  controlled by $\Delta$, 
acts only when two particles are at the closest possible positions 
$(i,i+t+1)$. The case $t=0$ recovers the standard XXZ chain (2) or (1).
In terms of Pauli matrices we can also write the equivalent Hamiltonian
\bea
H_{t}(\Delta) = -\frac{1}{2}\sum_{i=1}^LP_t(\sigma_i^x\sigma_{i+1}^x +
\sigma_i^y\sigma_{i+1}^y + \Delta\sigma_i^z\sigma_{i+t+1}^z)P_t \,,
\eea
where now $P_t$ projects out from the Hilbert space the states 
(in the $\sigma^z$-basis) where two up spins $(\sigma_t =1)$ are at distance 
smaller or equal to $t$. In coming from (1) to (2) or from (3) to (4)
we neglect a term proportional to a magnetic field. This term can be 
trivially added in the above Hamiltonian since the magnetization 
$\sum_{i=1}^L \sigma_i^z$ or the total number of particles
$n = \sum_{i=1}^zn_i$ are good quantum numbers. In terms of Pauli 
matrices the operator $P_0 = 1$ and for $t>0$ it is non-local and can 
be written in the $\sigma^z$-basis as
\bea
P_t = \prod_i \left[\ffrac{1}{2}(1-\sigma_i^z) + \ffrac{1}{2}(1 + 
\sigma_i^z) \prod_{l=1}^t \ffrac{1}{2}(1 -\sigma_{i+l}^z) \right]\,.
\eea
The Hamiltonians (2), although looking artificial, may describe interesting 
physical situations where the states with up spins are associated with larger 
excluded volume than those of  down spins. This is precisely the 
case when we consider  the time fluctuations due to diffusion 
of molecules with size $t+1$ (in units of the lattice spacing parameter) 
on a lattice. Interpreting  the master equation associated to this
problem as a Schr\"odinger equation in Euclidean time,\,\cite{2} the 
associated Hamiltonian is given by the isotropic ferromagnetic version 
of (3), namely $H_t(\Delta = +1)$. The anisotropic ferromagnetic version 
$(\Delta>1)$ is related to the asymmetric diffusion of the molecules of 
size $(t+1)$.\,\cite{3}

\section{The Bethe ansatz equations}

It is important to observe that similarly to the XXZ chain the 
$z$-magnetization $m = \sum_i^L\sigma_i^z$, or the total number of up 
spins $n$ (or the number of particles in (2) and (3)) is a conserved
quantity for the Hamiltonian (4). Consequently the Hilbert space can be 
separated into block-disjoint sectors labelled by $n =0, 1, \ldots, L/(t+1)$. 
In a given sector $n$ the wave functions of (4) can be written 
\bea
\Psi = \sum_{x_{1}=1}^{\bar L}\ldots
\sum_{x_{i+1}=x_i +t+1}^{\bar L +i(t+1)}\ldots\sum_{x_n=x_{n-1}+t+1}^{ L}
a(x_1,x_2,\dots,x_n)|x_1,x_2,\dots,x_n \rangle\,,
\eea
where $\bar{L} = L-n+1-t(n-1)$ and $a(x_1,...,x_n)$ are the amplitudes
of the spin configurations $|x_1,x_2,...,x_n\rangle$, having spin down at
the coordinates $x_i (i=1,2,...,n)$. We can calculate the above 
amplitudes and the corresponding eigenenergy of (4) by  applying the
 Bethe ansatz
\bea
a(x_1,x_2,...,x_n) = \sum_{P} A_{k_{p_1},k_{p_2},...,k_{p_n}}
e^{i(k_{p_1}x_1 + ... + k_{p_n}x_n)} \,,
\eea
where the summation is over all permutations $P = (p_1,p_2,...,p_n)$
and $A_{\{k\}}$ as well  the wave numbers $\{k\} = {k_1,k_2,...,k_n}$ 
should be calculated from the eigenvalue equation for (4). The solution
follows the standard coordinate Bethe ansatz.\,\cite{4} 
The validity of (7) for the case $n=1$ is immediate due to the 
translational invariance of (4), and in
this case the energies are $E_k = -2\cos k - \frac{\Delta}{2}(L-4)$,
with $k = 2\pi p/L$ $(p =0,1,2,...,L-1)$. The case of general $n$ can be 
verified as in the standard XXZ chain $(t=0)$. The energies and 
momentum are given by
\bea
E_{{k}} = -2\sum_{i=1}^n \cos 2k_i - \frac{\Delta}{2}(L - 4n)\,, \quad 
          P = \sum_{i=1}^n k_i \,.
\eea
The wave-numbers $\{k_i\}$ are obtained by solving the Bethe ansatz 
equations 
\bea
e^{ik_j L} = (-1)^{n-1}\prod_{l=1}^n e^{it(k_j-k_l)}
\frac{1-2\Delta e^{ik_j}+e^{i(k_j+k_l)}}{1-2\Delta e^{ik_l}+e^{i(k_j+k_l)}}\,.
\eea
These equations at $t=0$ give us the well-known Bethe ansatz equations  
for the XXZ chain. For $t>0$ these equations in a given sector 
specified by $n$ up spins and momentum $P=\frac{2\pi}{L}p$ $(p=0,1,...,L-1)$
are the same as those of an XXZ chain also with $n$ up spins  but 
lattice size $L'=L-tn$, momentum $P' = \frac {2\pi}{L-tn}p$ 
$(\mbox{mod} 2\pi)$ and twisted boundary condition\,\cite{5}
\bea
\sigma_{L+1}^x\pm i\sigma_{L+1}^y = e^{i\Phi}(\sigma_1^x \pm i\sigma_1^y)\,,
\quad \sigma_{L+1}^z = \sigma_1^z\,, \quad \Phi = \frac{2\pi}{L}p\,.
\eea
We will now study the phase diagram of these models. Calculating the 
eigenspectra of (4) numerically by 
brute-force diagonalization on small lattices, we see that in 
absence of an external magnetic field the groundstate, for finite 
chains will belong to  a sector where the magnetization $M=2n-L$
depends on the lattice size $L$, exclusion parameter $t$ and 
interaction $\Delta$, as long as $t>0$. This is distinct from the 
standard XXZ case  $(t=0)$, where the groundstate is always in
the half-filled sector $n=\frac{L}{2}$, with zero magnetization 
for all values of $L$ and $\Delta < 1$. In order to have a well defined  
thermodynamic limit for given values of $t$ and $\Delta$ it is
necessary to work with fixed values of  the magnetization per site 
$m=\frac{M}{L}=\frac{2n}{L}-1$, or equivalently for fixed values of 
the density of up spins $\rho =\frac{n}{L}$. 
We can force the groundstate to belong to such a sector with 
$n=\rho L$ by adding in (4) an external magnetic field
$-h\sum_{i=1}^{L}\sigma_i^z$, which will depend on the values of 
$t$, $\Delta$ and $\rho$.

It is important to remark that even when $\Delta =0$ we do not 
have a completely free system for $t>0$, since in (9) a phase
connecting different values of ${k_j}$ is always present. 
Nevertheless the Bethe-ansatz equations can be solved exactly 
for finite lattices, and is then instructive to calculate the 
eigenspectra at $\Delta = 0$.  

\section{The operator content at $\Delta=0$}

Let us consider $H_t(\Delta =0,h)$, with a given value of the magnetic 
field $h=h(\rho)$ that produces a fixed density 
$\rho =\frac{n}{L}$ $(0\leq\rho\leq\frac{1}{t+1})$ of up spins.
Taking the logarithm of both sides of (9) we obtain
\bea
Lk_j = \pi(\rho L-1) + \sum_{l=1}^{\rho L}t(k_j-k_l) +  2\pi m_j\,,\quad 
j =1,2,\ldots \,,
\eea
where $m_j$ $(j=1,2,...,n)$ are integers. The solution of these 
equations are given by the set $\{k_j\}$, 
\bea
k_j = \Theta + \frac{2\pi}{L(1-t\rho)}m_j\,, \quad 
\Theta = \frac{\pi(\rho L- 1)-tP}{L(1-\rho t)}\,, \quad j = 1,2,\ldots,n\,,
\eea
where $\{m_j\}$ are distinct integers and $P$ is the momentum 
associated to the state
\bea
P = \sum_{j=1}^{\rho L}k_j = \frac{2\pi}{L}\tilde p \; (\mbox{mod} 2\pi)\,, 
\quad \tilde p \in Z \,.
\eea
The groundstate has zero momentum and corresponds to the set $\{k_j\}$,
where
\bea
k_j = \frac{2\pi}{L(1-t\rho)}j\,, \quad j = -\frac{\rho L -1}{2}, 
-\frac{\rho L -1}{2} +1, \ldots ,\frac{\rho L -1}{2} \,.
\eea
Using (14) in (8) we obtain the groundstate energy of (4), in 
the presence of an external magnetic field, 
\bea
E_0(L,\rho) = -2\sin\left(\frac{\pi\rho}{1-t\rho}\right)
/
\sin\left(\frac{\pi}{L(1-t\rho)}\right) - h(2\rho-1)L\,.
\eea
The magnetic field $h=h(\rho)$ is to be calculated by imposing
that when $L \to \infty$, $e_{\infty} = \frac{E_0(L,\rho)}{L}$ 
will be a minimum as a function of $\rho$, i.e.
\bea
h =h(\rho) = \frac{t}{\pi}\sin\left(\frac{\pi\rho}{1-t\rho}\right)-
\frac{1}{1-t\rho}\cos\left(\frac{\pi \rho}{1-t\rho}\right) \,.
\eea
This result also tells us that in the absence of a magnetic field the 
groundstate should belong to the sector with magnetization 
$m_0 = \frac{M_0}{L} = 2\rho_0 - 1$, where $h(\rho_0) = 0$.
In the case $t=0$, $\rho_0 =\frac{1}{2}$ and $m_0 =0$. For $t>0$
we should solve $h(\rho_0) = 0$ numerically; for example, 
$\rho_0 = 0.3008443$, $\rho_0 = 0.2193388$ and $\rho_0 = 0.07354512$   
for $t=1,2$ and $10$, respectively.
As we can see these values cannot be predicted by symmetry 
arguments on (4) or (9).
The leading finite-size corrections of (15) are given by
\bea
\frac{E^0 (L,\rho)}{L} = e_{\infty} - \frac{\pi c}{6L^2}\xi
+ o(L^{-2})\,,
\eea
where
\bea
e_{\infty} = -2 \,\frac{1-\rho t}{\pi}\,\sin\left(\frac{\pi \rho}
{1 -\rho t}\right) - h(2\rho-1)\,, \quad 
c\,\xi = \frac{2}{1-t\rho}\sin\left(\frac{\pi\rho}{1-\rho t}\right)\,.
\eea
These relations are a clear indication that the model is critical 
for arbitrary values of $t$ and $\rho$, since (17) is the 
expected behavior of a massless conformally invariant critical 
point with $c$ and $\xi$ being the conformal anomaly and sound 
velocity, respectively.\,\cite{6} The sound velocity can be 
calculated from the energy momentum dispersion relation. 
The lowest eigenenergy in the sector with $n = \rho L$ spins up, 
having momentum $\frac{2\pi}{L}\tilde p$ $(\tilde p = 1,2,...)$ can be 
obtained by choosing the particular solution of (11),
\bea
k_j &=& \frac{2\pi}{L(1-t\rho)} j - \frac {2\pi t}{(1-t\rho)L^2}\,,
\nonumber \\ 
  j &=& -\frac{\rho L-1}{2},  -\frac{\rho L-1}{2} + 1, \ldots,
\frac{\rho L-3}{2}, \frac{\rho L-1}{2} + \tilde p \,,
\eea
and the corresponding energy is given by
\bea
\lefteqn{
E^{(\tilde p)}(L,\rho) = E^0(L,\rho)\; +} \nonumber\\ 
&&4\,\sin\left(\frac{\pi(\rho L + \tilde p - 1)}{(1+t\rho)L}\right)
\sin\left(\frac{\pi\tilde p}{L(1-t\rho)}\right) + o(L^{-1}) \,.
\eea
This gives us, for large $L$, a linear dispersion relation
with sound velocity
\bea
\xi = \frac{2}{1-t\rho}\sin\left(\frac{\pi\rho}{1-t\rho}\right),
\eea
and consequently the conformal anomaly is $c=1$ for 
all values of $t$ and $\rho$.
The conformal invariance of the infinite critical system, 
beyong relation (17), also predicts the finite-size 
corrections of excited states.\,\cite{7} For each primary 
operator, with dimension $x_{\phi}$ and spin $s_{\phi}$,
in the operator algebra of the critical system there 
exists an infinite tower of eigenstates of the Hamiltonian, 
whose energy $E_{m,m'}^{\phi}$ and momentum $P_{m,m'}^{\phi}$, 
in a periodic chain are given by
\bea
E_{m,m'}^{\phi}(L) &=& E^0 + \frac{2\pi\xi}{L}(x_{\phi}+m+m')+
o(L^{-1})\,,\nonumber\\
P_{m,m'}^{\phi} &=& (s_{\phi}+m-m')\frac{2\pi}{L}\,,
\eea
where $m,m' = 0,1,2, \ldots $.
The groundstate corresponds to the identity operator 
$x_{I}=0, s_{I}=0$ and the tower of states 
$E_{m,m'}^{(\phi =I)} = \frac{2\pi}{L}(m-m')$
in our model are obtained by decreasing (increasing) the 
values of the quasi-momenta $k_i$ in the set $\{k_i\}$ given in
(14). After the calculation of the eigenenergies 
corresponding to many of these descendants we convince ourselves that  
 the degeneracy of such levels are given by 
$d_{m,m'}^{\phi =I} = P(m)P(m')$ where $P(m)$ is the
number of integer partitions of $m$.
Another conformal dimension in the sector $n=\rho L$
can be obtained by taking negative (positive) 
values of $k_j$ in the groundstate root 
configuration (14) and replacing them by positive 
(negative) values. For example, the lowest of these
eigenenergies corresponds to the solution
\bea
k_j &=& \frac{2\pi}{(1-\rho t)L}(j - \rho t)\,, \nonumber \\
j &=& -\frac{\rho L-1}{2} +1, -\frac{\rho L-1}{2} + 2,\ldots ,
\frac{\rho L-1}{2}+1\,,
\eea
and has the value
\bea
E = -2\cos\left(\frac{2\pi}{L}\right)\sin\left(\frac{\pi\rho}{1-\pi\rho}
\right)/ \sin\left(\frac{\pi}{(1-\rho t) L}\right)
\eea
and a macroscopic momentum $P=2\pi\rho$. Comparing  
the finite-size corrections of (24) with (22) we see
that this energy corresponds to an operator with dimension
$x_{0,1}$ and spin $s_{0,1}$ given by
\bea
x_{0,\bar m} = (1-\rho t)^2 {\bar m}^2\,, \quad 
s_{0,\bar m} = n\bar m = \rho L\bar m\,,
\eea
respectively. The dimensions $x_{0,\bar m}$, with 
$\bar m$ positive (negative) integer are obtained by taking 
$\bar m$ of the lowest (highest) values of $k_j$ in (14)
and replacing them by the lower (higher) available 
positive (negative) values. As in the 
case of the identity operator the descendants of the above 
operators, with dimension $x_{\bar m}+m+m'$ $(m,m'\in Z)$
are obtained by increasing (decreasing) the positive
(negative) values of $\{k_j\}$ in the configuration that produces
$x_{0,\bar m}$. As before we convince ourselves that the degeneracy 
of these states is $P(m) P(m')$.

     There exist also dimensions associated to eigenstates produced by the 
addition $(\bar n = -1,-2,\ldots)$ or subtraction $(\bar n =1,2,\ldots)$ 
of spins up in the groundstate 
sector. In this case $n = \rho L + \bar n$ and the lowest eigenenergy 
is obtained from the set $\{k_j\}$ with
\bea
k_j = \frac{2\pi}{L(1-\rho t -\bar n)}j\,, \quad j = -j_{\rm max},
-j_{\rm max}+1,\ldots, j_{\rm max}-1, j_{\rm max}\,, 
\eea
where $j_{\rm max}=\frac{\rho L +\bar n -1}{2}$. These are 
zero-momentum eigenstates with energies
\bea
E^{\bar n} &=& -\;2\sin\left(\frac{\pi(\rho L +\bar n)}
                              {L(1-\rho t) -\bar n}\right)/
\sin\left(\frac {\pi}{L(1-\rho t) -\bar n}\right)\nonumber\\
&\phantom{=}& -\; h(2(\rho L+\bar n)-L)\,.
\eea
The finite-size corrections, when compared with (22) give us
the conformal dimensions
\bea
x_{\bar n, 0} = \frac{\bar n^2}{4(1-t\rho)^2}\,.
\eea

In the general case, where we combine $\{k_j\}$ configurations 
which give (25) and (26) we obtain the conformal dimensions
\bea
x_{\bar n,\bar m} = \bar n^2 X_p + \frac{\bar m^2}{4X_p}\,,\quad
X_p = \frac{1}{4(1-t\rho)^2}\,\quad  \bar n, \bar m \in Z
\eea
of primary operators with spin $s_{\bar n,\bar m} = 
(\rho L + \bar n)\bar m$. If we calculate the eigenenergies 
which correspond to the descendants with dimensions
$x_{\bar n,\bar m}^{m,m'} = x_{\bar n,\bar m}+m+m'$ we 
obtain their degeneracy $ d_{\bar n,\bar m}^{m,m'} =
P(m) P(m')$. The dimensions (29) are the same as the dimensions
$x_{\bar n,\bar m}^G = \frac{\bar n^2}{4\pi K} + \pi K\bar m^2$
of operators with vorticity $\bar m$ and spin-wave number $\bar n$
in a Gaussian model\,\cite{8} with coupling constant 
$K = (1-\rho t)^2/\pi$. The degeneracy $d_{\bar n,\bar m}^{m,m'}$
imply that the operator content of $H_t(\Delta = 0)$ is given
in terms of U(1) Kac-Moody characters. The situation is completely
analogous with that of the standard XXZ chain $(t=0)$ in its
massless regime.\,\cite{9} However, expression (29) shows different
dependences on the density $\rho$  (or magnetization),
depending on the value of $t$. In particular for
$\Delta = 0$ the dimensions are $\rho$-independent only in the 
free-fermion case $t=\Delta = 0$.
The result (29) shows that the net effect of the parameter $t >0$ at 
$\Delta =0$ is similar to that of the anisotropy parameter $\Delta \neq 0$ 
in the standard XXZ chain $(t = 0)$. In both cases the exponent $X_p$ changes 
continuously with the density $\rho$ of up spins. 
It is interesting to remark that when $\rho \rightarrow 0$ $X_p = 1/4$, for 
all values of $t$. This independence ot $t$ arises since the restrictions 
imposed by $t >0$ are not important due to the small amount of up spins in 
this limit. The other limit of $X_p$ at high density,
$\rho \rightarrow 1/(t+1)$, can also be understood as follows. If we start 
with the maximum density $\rho = 1/(t+1)$ and we take out one spin up, 
the ``hole" produced will have the same effective 
motion as that produced by adding $t+1$ ``particles" in the low-density 
limit. Then according to (28) we should obtain the limiting value
 $X_p = (t+1)^2/4$.

\section{The operator content at $\Delta \neq 0$}

    Let us consider the case where $\Delta\ne0$. The Bethe ansatz 
equations (9) supplemented by numerical calculations give the
same phase diagram for  all values of $t$. For $\Delta>1$
the groundstate is $(t+2)$-degenerate, where $(t+1)$ states
are in the sector with maximum number of up spins
$(n = \frac{L}{1+t})$ and the last in the sector with no spins 
up $(n=0)$. For $\Delta < 1$ we expect a massless behavior
similar to that in the XXZ chain. In fact, exploring the 
similarity of our equations for arbitrary $t$, with those of
the XXZ chain $(t=0)$, we can apply the method of Ref.~10
to extract the finite-size corrections for arbitrary values of 
$\Delta$ and $\rho$. The dimensions we obtained are given by
(29), where $X_p$ is now given by
\bea
X_p = \frac{1}{4}(1-t\rho)^{-2}{\eta }^{-2}(U_0)\,,
\eea
where $U_0$ and $\eta(U)$ are obtained by solving some integral
equations. These equations for $-1\leq \Delta=-\cos\gamma \leq 1$ are
given by
\bea
Q(U) &=& \frac{1}{2\pi}\frac{\sin\gamma}{\cos U-\cos\gamma}
-\frac{1}{2\pi}\int_{-U_0}^{U_0} 
\frac{\sin(2\gamma) Q(U')}{\cosh(U-U')-\cos(2\gamma)}dU'\,,  \\
1 &=& \eta(U) + \frac{1}{2\pi}\int_{-U_0}^{U_0}
\frac{\sin (2\gamma)\eta(U')}{\cosh(U-U')-\cos (2\gamma)}dU'\,,
\eea
while for $\Delta = -\cosh\gamma < -1$ they are given by
\bea
Q(U) &=& \frac{1}{2\pi}\frac{\sinh\lambda}{\cosh\lambda - \cos U}
-\frac{1}{2\pi}\int_{-U_0}^{U_0}
\frac{\sinh (2\lambda) Q(U')}{\cosh (2\lambda) -\cos(U-U')}dU'\,,  \\
1 &=& \eta(U) + \frac{1}{2\pi}\int_{-U_0}^{U_0}
\frac{\sinh (2\lambda) \eta(U')}{\cosh (2\lambda) -\cos(U-U')}dU'\,.
\eea
In Eqs.~(31)-(34) $U_0$ is calculated by imposing
\bea
\int_{-U_0}^{U_0}Q(U) =  \cases{
\frac{\rho}{1-t\rho}, & $0 \leq \rho \leq \frac{1}{2+\rho}$\,, \cr
\frac{1-(t+1)\rho}{1-t\rho}, &$\frac{1}{2+\rho}\leq \rho \leq 
\frac{1}{1+\rho}$\,.\cr}
\eea
The value of $X_p$ is obtained by numerically solving Eqs.~(31)-(35).
In Fig.~1 we plot $X_p$ at $\Delta = -\frac{\sqrt2}{2}$ for 
$t = 0,1,2$ and  3, while in Fig.~2 $X_p$ is shown for $t =2$ and some 
values of $\Delta$ $(\-1 \leq \Delta \leq 1)$.
 
\begin{figure}[htbp]
\begin {center}    
\mbox{\psfig{figure=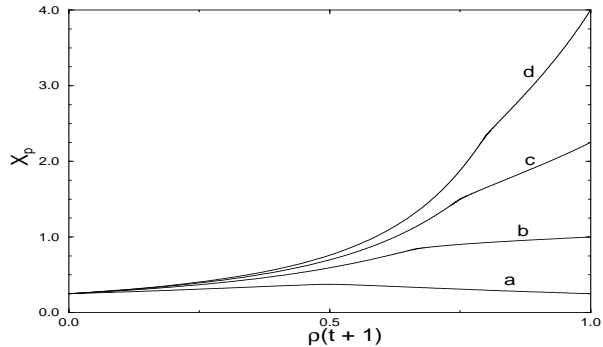,height=2.0in,width=2.0in,angle=-90}}
\end{center}
\caption{
The value of $X_p$ as a function of $\rho =\frac{n}{L}$ at
$\Delta = -\frac{\sqrt 2}{2}$, for $t=0$ (a), $t=1$ (b), $t=2$ (c)
and $t=3$ (d). }    
\label{FIG1}   
\end{figure}
 
\begin{figure}[htbp]
\begin {center}
\mbox{\psfig{figure=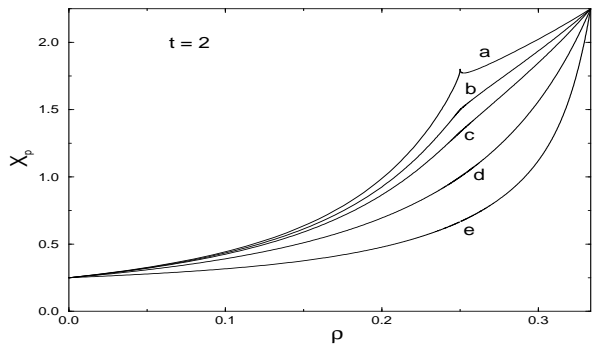,height=2.0in,width=2.0in,angle=-90}}
\end{center}
\caption{The value of $X_p$ as a function of $\rho =\frac{n}{L}$ for $t=2$
and anisotropy $-1 \leq \Delta = -\cos \gamma \leq 1$, with
(a) $\gamma = \frac{\pi}{10}$, (b) $\gamma=\frac{\pi}{4}$, (c)
$\gamma = \frac{\pi}{3}, (d) \gamma = \frac{\pi}{2}$ and (e)
$\gamma = \frac{2\pi}{3}$.}
\label{FIG2}
\end{figure}

These results coincide with those calculated earlier for the standard 
XXZ chain $(t =0)$.\,\cite{10}  At low density $(\rho \rightarrow 0)$ 
and high density $(\rho (t + 1) \rightarrow 1)$ the exponent 
$X_p = 1/4$ and $X_p = (t + 1)^2/4$, respectively.  These results are 
the same as those of the $\Delta =0$ case, since in these limits the 
distribution of ``particles" at $\rho \rightarrow 0$ or ``holes" 
at $\rho \rightarrow 1/(t + 1)$ are so diluted as to render the effect 
of the interaction term $\Delta$ negligible. From (30)-(34) it is 
possible to relate $X_p(t,\Delta)$ for different values of $t$ and $\rho$. 
In particular at the point $\rho = 1/(2 + t)$, which corresponds to the 
``half-filled" density for these models, we can show that 
\begin{equation}
x_p(t,\frac{\rho}{2+t}) = \ffrac{1}{4}(t+2)^2 X_p(t=0,\rho=\ffrac{1}{2}) = 
\ffrac{1}{8}(t+2)^2(\pi -\gamma)\,, 
\end{equation}
for $-1 \leq \Delta =-\cos{\gamma} \leq 1$, where we have used the known 
results\,\cite{5} for the XXZ chain in the absence of an external 
magnetic field.

In Fig.~3 we plot $X_p$ for $t = 3$ and for some values of $\Delta \leq -1$, 
while in Fig.~4 $X_p$ is shown at $\Delta = -2$ for $t =0,1,2$ and 3. 
 
\begin{figure}[htbp]
\begin {center}
\mbox{\psfig{figure=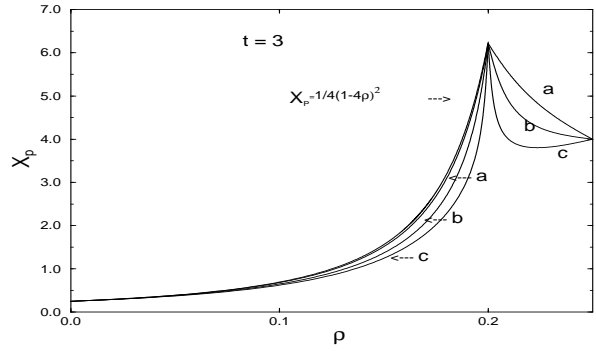,height=2.0in,width=2.0in,angle=-90}}
\end{center}
\caption{The curves $X_p(\rho)$ for $t=3$ and (a) $\Delta = -20$, (b) 
$\Delta = -4$, (c) $\Delta = -2$. The expected curve at 
$\Delta \rightarrow -\infty$ $X_p = 1/[4(1-4\rho)^2]$ is also shown.}
\label{FIG3}
\end{figure}

\begin{figure}[htbp]
\begin {center}
\mbox{\psfig{figure=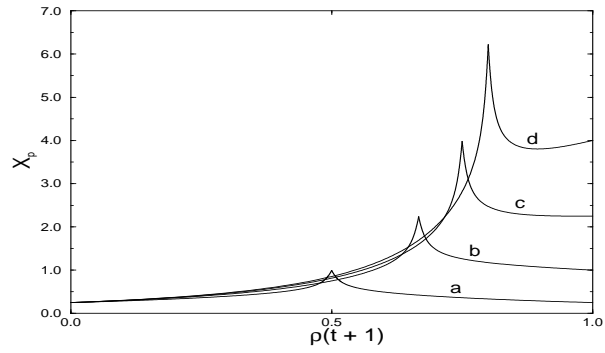,height=2.0in,width=2.0in,angle=-90}}
\end{center}
\caption{The curves  $X_p(\rho)$ at $\Delta = -2$ for (a) $t=0$, (b) $t=1$,
(c) $t=2$ and (d) $t=3$.}
\label{FIG4}
\end{figure}

As before in the low density $(\rho \rightarrow 0)$ and high density $(\rho 
\rightarrow 1/(t+1))$ limits we recover the $\Delta =0$ case. In Fig.~3 we 
also notice that at the ``half-filling" density $\rho = 1/(2+t)$ the 
exponent $X_p=(t+2)^2/4$, independently of the value of $\Delta$. A physical 
explanation for this behavior is the following. Strictly speaking at  the 
``half-filling" point $\rho = 1/(t+2)$ the model has a gap for $\Delta < -1$, 
as long as the magnetic field is zero. When the magnetic field increases, 
reaching a critical value $h = h_c(\Delta,t)$ this gap vanishes. Therefore 
the main physical effect at $\rho = 1/(2+t)$ is just the vanishing of the gap 
and not the interaction $\Delta$ 
itself. In the limit $\Delta \rightarrow -\infty$ 
and $\rho < 1/(2+t)$ the model with a given value of $t$ behaves effectively 
as a model with $t' = t+1$ and $\Delta =0$, since the spins up have infinite 
repulsion and will be located in positions with no interactions among them 
(distances greater than $t+1$). This can also be verifyed directly by setting 
$\Delta \rightarrow -\infty$ in the Bethe ansatz equations (9). 
Consequently for $\Delta \rightarrow -\infty$ and $\rho < 1/(2+t)$ we should 
have $X_p = 1/[4(1-\rho(t+1))^2]$. In particular at the ``half-filling" 
density $\rho = 1/(2+t)$ we obtain the value $X_p = (t+2)^2/4$. In Fig.~3 
the limiting curve $(\Delta \rightarrow -\infty)$, $X_p = 1/4(1-4\rho)^2$, 
for $t=3$, can be compared with that of $\Delta =-20$, obtained by solving \
(30) and (33)-(35).  For $\Delta \rightarrow -\infty$, but $\rho >1/(2+t)$ 
this argument does not apply and we should recover the value $X_p = (t+1)^2/4$ 
at $\rho =1/(t+1)$. We believe the results reported here in the $\Delta <-1$ 
region are unknown even for the standard XXZ chain $(t = 0)$.

\section{Generalizations}

    Before concludiing let us discuss an extension
of our results. We only considered the cases where $t\geq 0$
but what should be the quantum chain that would correspond 
to $t=-1$? In terms of particles, like in (2) this 
corresponds to the situation where we can have an arbitrary 
number of particles on a site. In fact we were able to find an 
exactly integrable quantum chain in this class, namely
\bea
H_{-1}(\Delta) &=& -\sum_{j=1}^L\{\sum_{n=0}^{\infty}
\sum_{m=n+1}^{\infty}(E_j^{n,m}E_{j+1}^{m,n} + 
E_j^{m,n}E_{j+1}^{n,m})\nonumber \\ & &+ \sum_{n=0}^{\infty}\Delta
(n-1)E_j^n E_{j+1}^n\} \,,
\eea
where $E^{n,m}$ are $\infty$-dimensional matrices located at 
the lattice sites with elements $E^{n,m})_{i,j} = 
\delta_{n,i}\delta_{m,j}$. The Bethe ansatz equations 
are given by (9) with $t=-1$. This
Hamiltonian commutes with the operator

\bea
\hat N =\sum_{j=1}^L\sum_{n=1}^{\infty}nE_j^{n,n}\,,
\eea
which gives the total number of particles. The Hamiltonian (37) 
in a given sector with $N$ particles can be written in terms of 
spin-$S$ SU(2) matrices with $S=(N-1)/2$.

\section*{Acknowledgments}

This work was supported in part by CNPq and FINEP -- Brazil, 
and also by the Russian Foundation of  Fundamental 
Investigations under Grant No.RFFI 97--02--16146.

\end{document}